\documentclass[a4paper,UKenglish]{lipics}
 
\usepackage{microtype}
\graphicspath{ {images/} }
\title{An Introductory Course in Logic for Philosophy Students: Natural Deduction and Philosophical Argumentation\footnote{This work was partially supported by UBACyT grant 01/Q532.}}
\titlerunning{An Introductory Course in Logic for Philosophy Students} 

\author[1]{Carlos A. Oller}
\author[2]{Ana Couló}
\affil[1]{Departamento de Filosofía,Facultad de Filosofía y Letras, Universidad de Buenos Aires \& Facultad de Humanidades y Ciencias de la Educación – IdIHCS, Universidad Nacional de la Plata\\
Argentina\\
  \texttt{carlos.a.oller@gmail.com}}
\affil[2]{Departamento de Filosofía, Facultad de Filosofía y Letras, Universidad de Buenos Aires\\
  Argentina\\
  \texttt{anacoulo@yahoo.com.ar}}
\authorrunning{C.\,A. Oller and A. Couló} 

\Copyright{Carlos A. Oller and Ana Couló}


\serieslogo{logo_ttl}
\volumeinfo
  {M. Antonia {Huertas}, Jo\~ao {Marcos}, Mar\'ia {Manzano}, Sophie {Pinchinat}, \\
  Fran\c{c}ois {Schwarzentruber}}
  {5}
  {4th International Conference on Tools for Teaching Logic}
  {1}
  {1}
  {169}
\EventShortName{TTL2015}

\begin{document}

\maketitle

\begin{abstract}
This paper tries to justify the relevance of an introductory course in Mathematical Logic in the Philosophy curriculum for analyzing philosophical arguments in natural language. It is argued that the representation of the structure of natural language arguments in Freeman's diagramming system can provide an intuitive foundation for the inferential processes involved in the use of First Order Logic natural deduction rules.
\end{abstract}

\section{Introdution}
In the 20th century Logic diverged from other philosophical fields by becoming a mature
formal discipline. By reaching this stage, Logic, under the form of Mathematical Logic, was able
to break away from its ancient interest in natural language arguments and its evaluation.
It has been said that the mathematization of Logic implied not just a change in methods,
but in the object of study as well. Furthermore, this entailed a pedagogical outcome: it became possible to design a good course in Elementary Formal Logic without ever mentioning the word “argument” in an ordinary sense. A great many Elementary
Formal Logic courses consist solely or mainly of First Order Logic and its metatheory,
that is, of a predominantly mathematical subject. Consequently, a significant number of
philosophy students feel that the compulsory taking of such a course is not only dull and
difficult but utterly irrelevant regarding their philosophic training (especially in those countries where analytic philosophy is not the predominant philosophical tradition). However, these complaints are frequently considered out of place, and moreover many logic teachers ignore, do not take into
consideration or disregard every pertinent criticism coming from the contemporary fields of Informal Logic or Argumentation Theory \cite{JohnBlair}. FOL has
been shown to be both insufficient and inadequate to address the whole wide field of
argumentation, especially philosophical argumentation. Besides, students’ protests
become reinforced by the current philosophical atmosphere, that has been developing a certain mistrust in formal Logic as a trustworthy guide for setting philosophy “on the sure path of science”. \par
In this paper we suggest a way of integrating the study of natural language
arguments with FOL in introductory logic courses for the Humanities (especially
Philosophy). We propose, firstly, that the relationship between argumentative strategies
that are identifiable in natural language arguments and natural deduction rules of
inference for FOL should be explicitly addressed. In addition to this we recommend
using a diagramming system for argument structure representation as a means to
offer an intuitive foundation for the inferential processes involved in using those
rules. These steps imply taking into consideration the aforementioned criticisms, and at
the same time, offering a plausible justification for the presence of FOL as an
introductory level mandatory subject in Philosophy departments.
\section{Reasons for teaching FOL as an introductory level mandatory subject to Philosophy students}
Several reasons might be given for the presence of FOL as part of the essential basic training for
philosophy students. A familiar one is the claim that studying FOL will develop
students’ abilities in identifying, reconstructing and analyzing natural language
arguments. However, both teachers’ experience and relevant research seem to disprove
this claim.\par
The underlying idea here is that the evaluation of a deductive argument expressed in a
natural language (such as Spanish, French or English) depends fundamentally on the
logical form of that argument when it is translated into the artificial language of a logical
system (such as FOL). This assumption frequently entails that even if an important part
of philosophers’ professional activity consists in building, analyzing and evaluating
arguments, the theories and procedures that underlay those tasks are seldom, if ever,
explicitly addressed in other subjects in philosophy departments and curricula. This job
is usually left in the hands of introductory logic courses, which consist mainly in an
introduction to deductive FOL. And, correlatively, introductory logic courses rarely take
the opportunity to integrate formal logic elements with philosophical problems at large.\par
However, classic research such as that of Cheng, Holyoak et al.\cite{Cheng}, brings attention, at the
very least, to the fact that explicit instruction in FOL does not necessarily entail an
amelioration of reasoning abilities: “Our results have clear educational implications.
We have shown that deductive reasoning is not likely to be improved by training on
standard logic.” Beebee \cite{Beebee} offers an interesting
example of the difficulty of applying formal logic as a means to elucidate the logic
behind a philosophical argument. She starts by asking about the relevant goals and
contents of a significant introductory formal logic course for a philosophy department.
She then sketches an exam item where she asked students “to identify the main rule of inference used in the following argument:
If God exists, he is omniscient, benevolent and all-powerful. There is evil in
the world. Suppose that God exists. Then, being omniscient, he knows there
is evil in the world. Being all-powerful, he could have prevented that
suffering. But he has not prevented it. This is incompatible with the claim
that he is benevolent. So God does not exist.”
She found out that only 10 to 15 percent of the students were able to identify \textsl{reductio ad absurdum} as the main rule of inference used in the argument, while a significant larger portion could
employ it adequately in a formal proof. But this suggests that one of the goals that
Beebee states for an introductory formal logic course for first year philosophy students
– understanding the logic behind philosophical arguments – cannot be reached just by
this means. Beebee’s example seems to complement the results reported by Cheng and
Holyoak: academic training in mathematical logic not only is not enough to enhance the
ability for deductive reasoning but it is also incapable by itself of enhancing the ability
to analyze deductive arguments expressed in natural language.\par
A second reason for the crucial question about the reasons that might be given for the presence of FOL and its metatheory as an essential part of the basic training for philosophy
students has to do with its relevance in enabling students to pursue further studies in
Logic, Epistemology or Philosophy of Science that would include advanced courses of
Logic as part of the curriculum. But this argument begs the question of the relevance of
Logic, and it does not take into consideration the diverging interests of students that
would choose other traditions or “streams”. Furthermore, Logic, Epistemology and
Philosophy of Science majors would certainly need to deepen their knowledge by taking
more than one course of Logic, so as to achieve a much deeper level of understanding of this discipline.\par
A compromising alternative to our question would be to save part of the course for an
informal introduction to logic that includes, as someone has quaintly put it “all those
issues that would never be part of a paper in the Journal of Symbolic Logic”. This
perspective tends to adopt a propaedeutic stance that addresses these issues rapidly and
superficially, so as to get to the really important questions. It is easily found in many introductory Logic texts, readings or handbooks for Humanities students. But sadly,
most texts do not integrate the “informal logic” and “formal logic” sections, and this can
bring about some problems. For instance, the central notions of “argument” and “logical form” that appear in the “informal logic” section usually differ from those appearing in
the “formal logic” one. The notion of “argument” in a natural language context – in which
philosophical arguments are enunciated- involves a pragmatic conception by reason of
which we recognize an argument by identifying in a text the intention to support of the
truth or the acceptability of a sentence –conclusion- by the truth of acceptability of other
sentence(s) (or argument(s)) –premises. On the other hand, the concept of argument
belonging to a formal logic approach is a mathematical notion, devoid of any pragmatic
underpinnings: an argument is an ordered pair whose first member is a (possibly empty) set of well-formed formulas (the premises), and whose second member is a well formed formula (the conclusion). Naturally, interpreting an argument in the second sense does
not necessarily produce an argument in the first sense, and the translation of an
argument expressed in natural language to a FOL language would be incapable of
conveying the pragmatic element that is essential for it in the first place.
\section{Integrating the teaching of natural language argumentative strategies and natural deduction rules of inference}
On the basis of the aforementioned discussion, our proposal for an introductory course
in Logic for philosophy students aims at weaving a tighter web between an informal
presentation of argumentative strategies typical of philosophical arguments – \textsl{reductio ad absurdum}, hypothetical reasoning, reasoning by cases, universal
instantiation, etc. – and the FOL rules that codify those strategies. In our experience,
familiarity with rules and proof building in FOL does not automatically mean
proficiency in detecting those inferential strategies that these rules and proofs translate
into philosophical arguments. On the contrary, it seems desirable to start by presenting
and examining those strategies by means of philosophical arguments in natural language
so as to help students understand the intuitive grounding of natural deduction rules that
will be presented in the course section set apart for FOL. Indeed, as we have argued,
when building proofs, most students tend to apply FOL natural deduction rules in a
mechanical way, without fully grasping their sense.\par
A good example of this difficulty in understanding the intuitive rules of inference
meaning can be found in the basic rule of conditional introduction. This rule involves
reasoning from assumptions or hypotheses and is an essential element in the
presentation of logic as a natural deduction system. Also, it is a frequent argumentative
strategy in philosophical argumentation. We can represent this logical rule in the usual way using the following diagram:

\includegraphics[width=5cm]{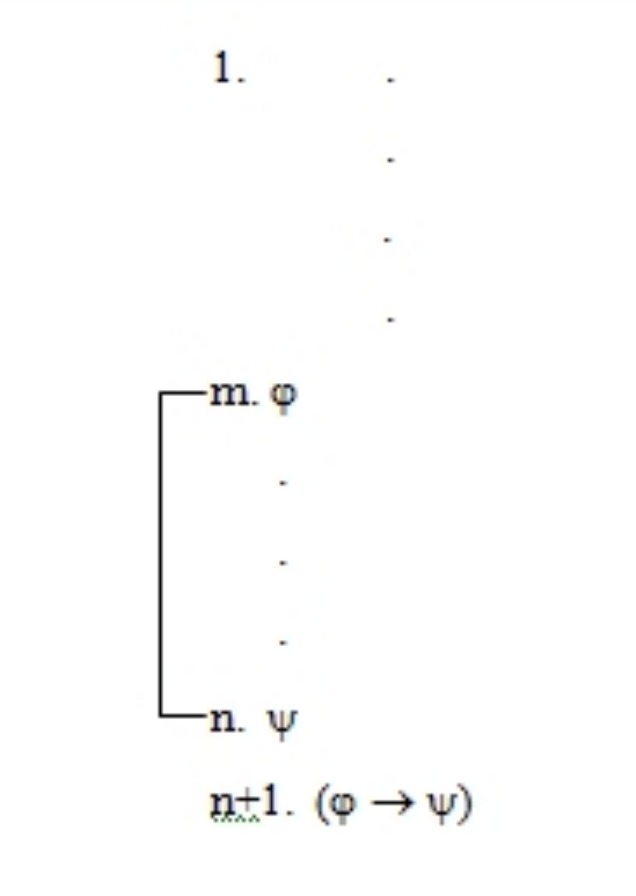}

Students usually learn quickly to build logical derivations that include this rule, by following the advice that recommends to assume the antecedent and try to derive the consequent, whenever one wants to derive a conditional. However, they frequently display difficulties when it comes to the point of having to understand the argumentative strategy that is formalized by that rule if it is embedded in philosophical
arguments such as the following: Ryle \cite{Ryle} affirms that the “intellectualist legend” states
that an intelligent act is intelligent so far as it is the result of an internal, previous plan that must be intelligent itself for the act to be so.At the same time, planning is a mental operation which, if it is to be deemed intelligent,must itself inherit that characteristic from a previous act of planning: the planning of
planning. This previous act, for its part, may be intelligent or stupid. But if it is intelligent,
it must itself inherit that characteristic from a previous act of planning, and so and so forth
world without end. Since we can find human beings that perform intelligent acts, we can
conclude that – under the assumption that the intellectualist legend is right – some human
beings are capable of performing an infinite number of mental acts. Therefore, if the
intellectualist thesis is true, there is (at least) one human being who is capable of performing
an infinite number of mental acts.
The difficulties that students display in understanding this argumentative strategy may be
due, at least in part, to the fact that in reasoning from assumptions (as we did in the
example) the conditional conclusion is supported by a complete (sub)argument rather than
by simple statements. But students do not expect this kind of support because it is not considered by the informal definition of argument that is usually given by the textbooks. However, this peculiarity that is
formally expressed in the rules that involve assumptions in the presentation of FOL as a natural deduction system – and that intend to reflect an argumentative strategy regularly
present in mathematical reasoning (and in other fields as well) – is rarely taught in basic
logic texts for the Humanities. Therefore, it is no wonder that many students can be
perfectly capable of mechanically applying the conditional introduction rule when building
a proof, and, at the same time, do not fully understand the strategy exemplified in Ryle’s argument.
\section{Argument diagramming and proofs in a natural deduction system}
We have found useful introducing the standard technique of argument diagramming
when trying to integrate the study of natural language arguments and the study of
arguments formulated in the mathematical logic languages. Argument diagramming
allows students to identify and represent the inferential relationships between the
sentences that constitute the arguments in natural language, without having to resort to
the formalization of those sentences in FOL language. In the philosophical tradition
pertaining to the analysis of argument structure, this technique can be found, for
instance, in James Freeman’s works \cite{Freeman1}\cite{Freeman2}.\par
Even if argument diagramming can be considered to be typical of informal logic strategies in the analysis of argument, it is also closely related to many issues linked to formal logic inferences.
In fact, the tree structure typical of the standard argument diagramming, allows students to understand the intuitive meaning of proof building in FOL. Most introductory logic textbooks present FOL proofs using Ja\'{s}kowski-Fitch style of natural deduction representation, a graphical method that presents proofs as linear sequences of formulas \cite{HazPel}.  But, Gentzen original presentation \cite{Szabo} conceived of them as finite trees: the root of the tree is the formula to be proved, the leaves of the tree are the assumptions and the other formulas are obtained by the application of an inference rule from the formulas standing immediately above it.\par 
The Gentzen representation of proofs allow us to display the logical support structure of arguments and to identify the subarguments of which complex arguments are built, i.e. what Freeman calls "the macrostructure of arguments". In this way, by identifying the argumentative strategies that natural deduction rules intend to codify, and by portraying derivations as special instances of Gentzen style diagrams, a reasoned and historically situated transition from arguments in natural language to mathematical logic derivations can be made possible.\par In order to illustrate this proposal, let us look at the following version of an argument presented by Plato in his Apology \cite{Plato}.
Death is one of two things: either death is a state of nothingness and utter unconsciousness,
or, as men say, there is a change and migration of the soul from this world to another. Now if you suppose that there is no consciousness, but a sleep like the sleep of him who is
undisturbed even by the sight of dreams, death is good. But if death is the journey to
another place, and there, as men say, all the dead are, then death is good. Therefore, in any
case, death is good.
This argument exemplifies the argumentative strategy of reasoning by cases and its standard diagram is the following:

\includegraphics[width=13cm]{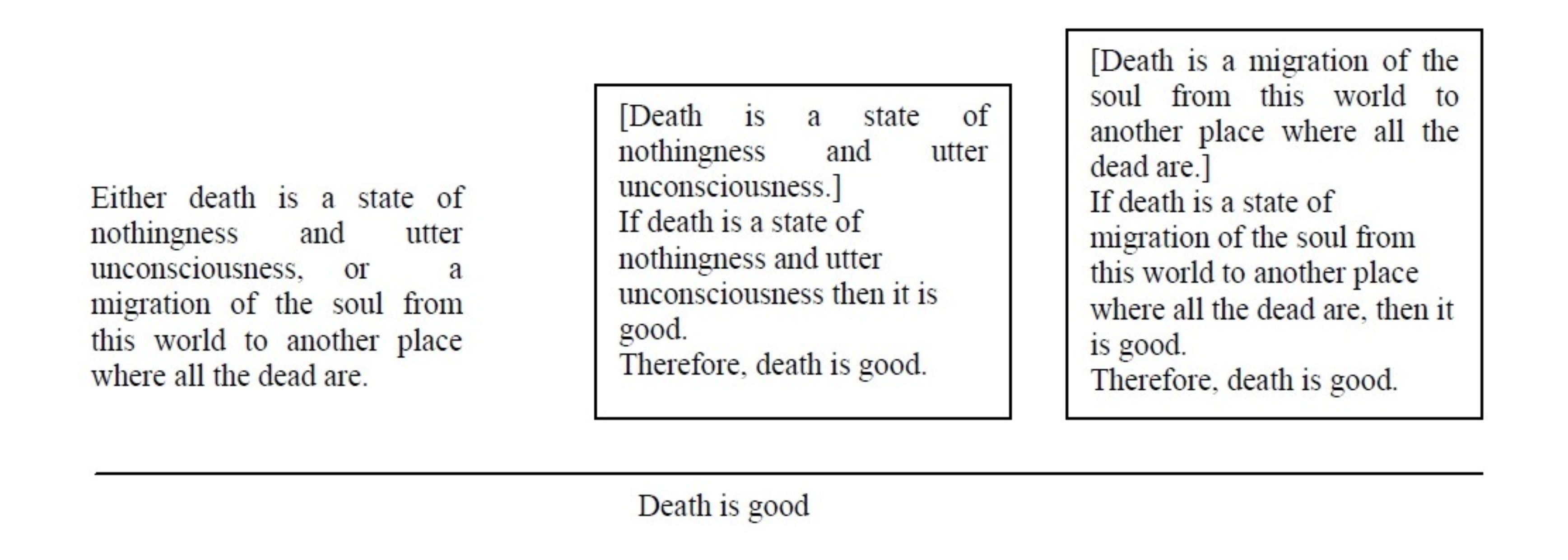}

The diagram – where the subarguments that support the conclusion are enclosed within a box, and assumptions are enclosed between brackets- makes evident that this argument is a case of reasoning from assumptions, and, in particular, an example of reasoning by cases. This strategy is represented by the rule of disjunction elimination, which in Gentzen’s natural deduction system adopts the following form:

\includegraphics[width=5cm]{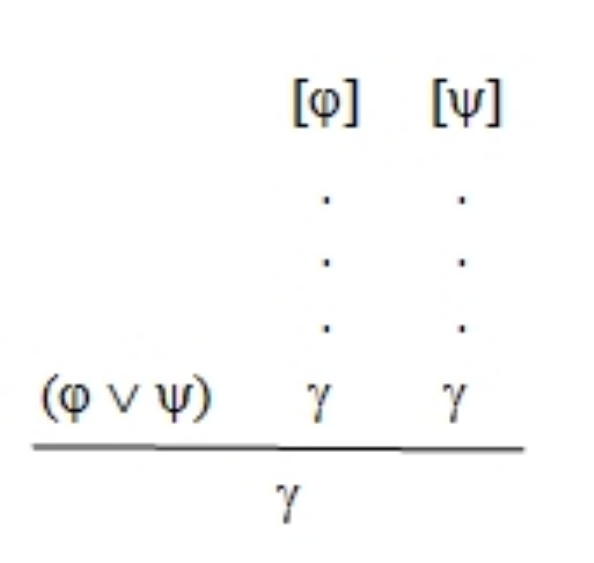}

The diagram that represents Plato’s argument as a tree whose conclusion is supported by linked premises clearly suggests the inferential strategy that can be applied to the FOL translation of the argument in order to derive its conclusion and provides an intuitive understanding of the inference rules involved in the derivation. In this way the close relation between the macrostructure of natural language arguments and derivations in First Order Logic is made evident.
\section{Conclusions}
 In this work we have presented a proposal that aims at the integration of natural deduction and philosophical argumentation in an introductory course of Logic for Philosophy students. We drew from, and conceptualized, the pedagogic experience obtained teaching the mandatory undergraduate Logic course offered by the Philosophy Department at the University of Buenos Aires.\par
On the one hand, we advised for the integration of the informal presentation of some argumentative strategies that are commonly found in philosophical argument with the FOL rules that codify those strategies in natural deduction systems.\par
On the other hand, we proposed that those courses incorporate the standard technique of argument diagramming that allow for the identifying and representation of the inferential relationships between sentences that are part of arguments in natural
language. These techniques offer students an opportunity to grasp the intuitive sense of
the building of proofs in FOL, and discover its relationship with the inferential structure
of arguments in natural language.\par
Based on these premises we aim at building a closer integration between the sections reserved for informal logic and those set apart for mathematical logic in introductory courses and textbooks of logic for the Humanities. It is to be hoped that this integration will bring to the fore the relevance of the mathematical logic content included in those courses for the study of natural language arguments, and especially for the study of philosophical argument.





\newpage
\thispagestyle{empty}
{\ }

\end{document}